\documentstyle[aps,prb,twocolumn]{revtex}

\begin{document}

\bibliographystyle{prsty}

\widetext

\title{Conductance of a quantum wire: Landauer's approach versus Kubo formula}
\author{In\`es Safi}
\address{Service de Physique de l'Etat Condens\'e, Centre d'Etudes de Saclay\\ 91191 Gif--sur--Yvette, France\\(received 22 December 96)}

\maketitle
\begin{abstract}
The transport in a pure one-dimensional quantum wire is investigated for any range of interactions.
First, the wire is connected to measuring leads. The transmission of an incident electron is found to be perfect, and the conductance is not renormalized by the interactions. Either Landauer's approach or Kubo formula can be used as long as the reservoirs impose the boundary conditions. Second, the Kubo formula as a response to the local field is reconsidered in a generic Luttinger liquid: the ``intrinsic'' conductance thus obtained is determined by the same combination of interaction parameters as that which renormalizes the current.
\end{abstract}
\draft


  \narrowtext

Any reliable study of a system of electrons should consider their
interactions.\ This challenging obstacle has
been successfully crossed in one dimension, giving rise to the so called
Tomonaga--Luttinger liquid (TLL).\cite{haldanebosonisation}\ The latter is characterized by collective modes
propagating with different spin and charge velocities.\cite{haldanebosonisation} In earlier works, the restriction to one dimension was viewed as a step towards understanding the physics of quasi-one-dimensional or higher dimensional systems. In particular, the conductance of a one-dimensional ballistic wire was computed as a formal quantity, and found to be renormalized by the interactions:\cite{apel_rice,group}
\begin{equation}
\label{conductanceK}g=\frac{2e^2}hK,
\end{equation}
where $K$ depends on the microscopic model, and $K=1$ in the absence of
interactions. Recently, it has become possible to fabricate ballistic quantum wires\cite{experiments} but
the prediction (\ref{conductanceK}) has not yet been observed. One has to reconsider the theory by taking the physical reality of these new systems into account, particularly the way the conductance is measured. A step was made towards this aim by connecting an
interacting wire to perfect noninteracting one-dimensional leads:\cite{ines,maslov} those are
intended to simulate the propagating mode through the two-dimensional Fermi gas
where the quantum wire opens.\ The role of the reservoirs is accounted for by the
flux they inject, in the spirit of Landauer's approach\cite{landauer:70}.\  In the presence of short range interactions, an
incident flux is perfectly transmitted.\cite{ines}\ According to Landauer-B\"uttiker's
formula relating the conductance to the transmission\cite{landauer:70,buttickermulti} (extended rigorously to the interacting wire\cite{ines}), the conductance of the wire is
\begin{equation}
\label{gnue}g=\frac{2e^2}h. 
\end{equation}
 
The purpose of this paper is twofold. First, we generalize the above result to long-range interactions by restricting screening to the interactions between electrons in the wire and those in the leads. The second part is a discussion of recent works where
the result (\ref{gnue}) is derived without reference to the measuring leads, either by following Landauer's spirit\cite{egger,shimizu} or using the Kubo formula.\cite{kawabata,fujimoto,oregrecent}\ Using the latter, we will show in a straightforward way that the conductance of a wire without reservoirs is actually given by

\begin{equation}
\label{glocal}g^{\prime }=\frac{2e^2}h\sqrt{\frac{uK}{v_F}}
,
\end{equation}
where $u$ denotes charge velocity. One can see that Eq. (\ref{glocal}) identifies with Eq. (\ref{gnue}) only in the case where $uK=v_F$, i.e., when the current is not renormalized by the interactions. Equation (\ref{glocal}) is more attractive than Eq.(\ref{conductanceK}) since it is independent on the interactions for systems with full (Galilean) translational invariance. Besides, contrary to Eq.(\ref
{conductanceK}), it conciliates the conductance of an
isolated wire with its sensitivity to boundary conditions expressed through
the charge stiffness $uK$.\cite{hubbard} This same combination will give the conductance if we adopt the hypothesis of Ref.\onlinecite{egger}.

A common useful tool to discuss those issues is the equation of motion. One can ignore spin, which can be accounted for by a factor $2$ in Eq.(\ref{gnue}) because the transport
 is determined by the charge degrees of freedom. Consider a general interaction Hamiltonian: $$H_{int}=\int dx dy U(x,y) \rho_{tot}(x)\rho_{tot}(y).$$ The density $\rho_{tot}$ has both a long wavelength part $\rho$ and a $2 k_F$ component.\cite{haldanebosonisation} The Hamiltonian can be cast in the quadratic form
\begin{equation}
\label{Hc}H=\int \frac{\pi dx}{2uK}\left[ j^2+u^2\rho ^2\right] +\int \int
U(x,y)\rho (x)\rho (y),
\end{equation}
provided $U$ varies smoothly on the scale $\pi/k_F$ and $\int dy U(x,y)(x-y)^n\cos (2k_Fy)=F_n(x)$ converges for any $n$.\cite{these} Then $u(x)$ and $K(x)$ are renormalized from their noninteracting value $(v_F,1)$ by the backscattering process\cite{these} $u(x)/K(x)=v_F+F_2(x)$. But $uK=v_F$ is not renormalized contrary to Refs.\onlinecite{haldanebosonisation,fabrizio_coulomb} and \onlinecite{schulz_wigner}. Nevertheless, we keep $uK$, since it can be renormalized by other irrelevant processes such as the umklapp process or electron-phonon coupling, etc.... If $U$ is short range, it can be absorbed in $u/K$, but could also affect $uK$ if interactions between electrons with the same or different velocities are distinguished. $j$ and $\rho$ can be expressed in term of the density for right and left moving electrons $\rho_{+,-}$ where $\rho=\rho_+ +\rho_-$ and $j=uK(\rho_+-\rho_-)$. They obey the commutation rule $\left[ j(x),\rho (y)\right]=i\delta'(x-y)$, where $\delta$ is the Dirac function, thus
\begin{equation}
\label{eqcompact}\frac 1{uK}\partial _tj+\partial _x\left( \frac{\partial H}{
\partial \rho }\right) =0
.
\end{equation}
\  Using the expression of $H$ in Eq.(\ref{Hc}) one gets:
\begin{equation}
\label{jrho}\frac {\partial _tj(x,t)}{uK}+\partial _x\left[ \frac uK\rho
(x,t)+\int dy U(x,y)\rho (y,t)\right]=0
\end{equation}
Finally, using the continuity equation, one can express this equation in term of $j$ or $\rho$ alone.

Before considering the effect of the external leads, let us recall briefly the
conductivity of an infinite wire with long range interactions, $U(x,y)=1/\left| x-y\right|$. The divergence at short separation is usually cutoff by a confinement length $l$.\ An external field forms a source term on the right--hand--side of Eq. (\ref{jrho}) which gives a straightforward way to recover the conductivity. The conductance decreases with the wire length $L$, 
\begin{equation}\label{gcoulomb}
g\sim 1/\sqrt{\log (L/l)} 
\end{equation}
up to some constants.\cite{fabrizio_coulomb,these} Suppose that the wire is now perfectly connected to noninteracting leads whose electrons do not interact with those in the wire. For instance, $U(x,y)=f(x)f(y)/\left| x-y\right|$, with $f$ decreasing smoothly to zero on the leads. 
Consider a right-going flux with density $\rho_+$ injected by
the left reservoir in the left lead. Choosing an arbitrary point $x_0$ on the latter and the time the flux reaches $x_0$ as the initial time, we have $\left\langle \rho
(x,t=0)\right\rangle =\left\langle j(x,t=0)\right\rangle /v_F=\rho_+\delta (x-x_0)$.
The stationary limit of Eq.(\ref{jrho}) once Fourier transformed, yields the uniformity of
\begin{equation}
\label{uniformdesity}\frac uK\rho (x,\omega =0)+\int dyU(x,y)\rho (y,\omega
=0)=v_F\rho_+
\end{equation}
The second term on the left hand side vanishes in the noninteracting leads, thus the
density is the same on both leads and is equal to $\rho_+$. Thus the transmission of an incident flux is {\em perfect}, as a consequence of the current uniformity. It has to be emphasized that no constraint on the wire length $L$ is needed, apart from the fact that the dc limit means $\omega L\ll 1$. Note that the zero mode corresponding to the number of electrons operates in Eq. (\ref{uniformdesity}): the system can exchange electrons with the reservoirs.

 We now consider the conductance. A first alternative is to continue with Landauer's approach, but this deals mainly with noninteracting systems.\cite{landauer:70} The transmission is a function of the energy that has to be conserved
during traversal.\ This is clearly not the
case in our interacting wire. Nevertheless, it helps us to know that the transmission is perfect.\ \ Each reservoir injects electrons at energies up to its electrochemical potential, giving a current
\begin{equation}
\label{jf}j_{+,-}=\frac{2e}h\int dEf_{L,R}(E) 
,
\end{equation}
where $R$ or $L$ denote the right or left reservoir. The cancelation between density and velocity in the one dimensional
leads is used to get Eq. (\ref{jf}). Of course, in a stationary regime, the electrons are continuously injected. Since the transmission is perfect, the net current is

\begin{equation}
\label{jmu}j=j_{+}-j_{-} =\frac{%
2e}h\left( \mu _R-\mu _L\right)=\frac{2e^2}h\left( V_R-V_L\right).
\end{equation}
 This result holds at any finite temperature much
less than $\mu _R$ and $\mu _L$. Thus the conductance is given by Eq.(\ref{gnue}) and is independent from interactions. The derivation of this result supposes implicitly that $\mu_{R,L}$ can be imposed independently on the current, and that the electrostatic potential $V_{R,L}$ (or more precisely $V_{loc}$ discussed later) varies as $\mu_{R,L}$ varies. In general, $V$ and $\mu$ are different:\cite{landauer:70} $V$ shifts the bottom band, and its variation generates an electric field.\ $\mu$ controls
the filling, and its variation generates the analogue of a diffusive force
that would act even if the electrons had no charge. $V$ and $\mu$ follow
each other wherever charge neutrality is ensured, and this is generally the
case in ``good'' reservoirs.\cite{landauer:70,buttickermulti} That is why we can also get a microscopic derivation of Landauer's formula%
\cite{ines}. The reservoirs can be modeled by the electrostatic potential $%
V_R,V_{L}$ they impose instead of the flux they inject.\ The latter
are determined by  $\mu _R,\mu _L$ [Eq.(\ref{jf})], but the stationary current in
the presence of an electric field depends also on $V_R-V_L=\mu _R-\mu _L$.\
The same Green's function associated with Eq. (\ref{jrho})
determines the time evolution for either a source term (the
electric field) or an initial condition (the injected flux).\cite{remarque} Then a dynamic relation between transport and transmission can also be
derived.\cite{ines} If another model is considered by including interactions in the leads with parameters $u_1,K_1$ different from $v_F,1$, such a relation has an additional factor $u_1K_1$. On the other hand, Eq. (\ref{jmu}) contains now the same prefactor due to the renormalization of the current in the leads. Thus both the microscopic and the phenomenological argument give the same conductance $u_1K_1/v_F$. But one has to suppose implicitly that the electrons injected by the reservoirs in interacting leads are not reflected back. This is not usually required: only a nonreflective absorption by the reservoirs is important. Thus a better description of the interface is needed if one considers interacting leads.

Let us now discuss other related works. A similar concept to the one we used following Landauer's spirit\cite{ines} was exploited in Ref.\onlinecite{egger}, but without including leads. This amounts to take $u_1=u$, $K_1=K$, thus the conductance is now $u_1K_1/v_F=uK/v_F$, generalizing Ref.\onlinecite{egger} to the case where the current is renormalized by the interactions. As discussed shortly before, this supposes the electrons are injected immediately in the wire without reflections. If one injects a right-going flux inside a wire with short-range interactions, the fraction $(1-K)/2$ is immediately reflected.\cite{these} The leads are often introduced because it would be difficult to define and to find the
transmission directly between real reservoirs.

Other works claimed to recover Eq. (\ref{gnue}) without using leads.\ In Refs.\onlinecite{shimizu,oreg}, the
potential for right- and left--going electrons was introduced as the
conjugate variable to their density

\begin{equation}
\label{muplusmoins}\mu _{+,-}=\frac{\partial H}{\partial \rho _{+,-}},
\end{equation}
thus $\mu_+-\mu_-=j$ using Eq. (\ref{Hc}). It is not clear why these nominal potentials should coincide with those
in the reservoirs that would have to accommodate the interactions in the wire.\ In Ref.\onlinecite{shimizu}, it was invoked that this must be so for
the stationary state to be stable. Nevertheless, the right (left) --going electrons through the
structure are not only those coming from the left (right) reservoir, but
also those generated by the reflections on the wire.\ It is not even
granted that they have any defined chemical potential or an equilibrium distribution.\ One has to prove the perfect transmission 
through the pure TLL for the potential of the right-- or left--going
carriers to coincide with these of the reservoirs.\ It was shown before\cite{ines} that the
change in interactions does not cause reflections.
Another series of works\cite{kawabata,oregrecent} also recovered the result (\ref{gnue})
without reference to the leads or to the reservoirs, but by computing the response to the local
electrostatic potential $V_{loc}$. In the second part we give our contribution to this problem, first on the technical level, than on the conceptual one. 

It was emphasized in Ref.\onlinecite{izuyama} that the Kubo formula has to be used with
respect to the local potential verifying:
$\Delta V_{loc}=\Delta V+\rho$, where $V$ is the external potential and $\Delta$ is the Laplacian in the three--dimensional space.\ If the wire is
isolated, the integration of this equation gives exactly a kernel formed by the long-range
interactions: $V_{loc}(r)=V(r)+\int\rho(y)/\left| r-y\right|$,
where the integral is restricted to the wire. In general, there are reservoirs and metallic gates
around, so that one has to solve an overall electrostatic problem.\ Note that this would enlighten us on the partially screened interactions $U(x,y)$ in
Eq. (\ref{Hc}). A general way to express the local potential is then
\begin{equation}
\label{Vloc}V_{loc}(x)=\frac{\partial H_{int}}{\partial \rho(x)},
\end{equation}
where $H_{int}$ is the interaction Hamiltonian, including external charges $\int \rho V$. It is easy to see that
\begin{equation}
\label{functional}\frac{\partial H_{int}}{\partial\rho}=\frac{\partial H}{%
\partial \rho }-v_F\rho,
\end{equation}
where the second term comes from the functional derivative of the kinetic
Hamiltonian. By the way, this expression shows the difference between the electrostatic potential (\ref{Vloc}) and the electrochemical potential (\ref{muplusmoins}) taken as
an average of $\mu_{+}$ and $\mu _{-}$: 
$$
\mu =\frac{\mu _{+}+\mu _{-}}2=\frac{\partial H}{\partial \rho}=V_{loc}+v_F\rho,
$$
 This illustrates the fact that $\mu$ and $V_{loc}$ deviate wherever charge
neutrality is broken.
Let us now compute the response to $V_{loc}$.\ Comparing the equations (\ref{functional}) and (\ref{eqcompact}), we see
immediately that the latter is equivalent to
\begin{equation}
\label{eqloc}\frac {\omega^2}{uK}j+v_F\partial _{xx}j =-i\omega E_{loc}, 
\end{equation}
where we used the continuity equation to eliminate the density.
\ This is an analogue to the equation of response to the external field in
a system with short--range interactions parametrized by $u^{\prime }$,$
K^{\prime }$:
\begin{equation}
\label{eqext}\frac {\omega^2}{u^{\prime }K^{\prime }}j+\frac{u^{\prime }}{%
K^{\prime }}\partial _{xx}j =-i\omega E 
\end{equation}
Comparing Eq. (\ref{eqloc}) with Eq. (\ref{eqext}), one gets
\begin{equation}
\label{uprimeKprime}
K^{\prime }= \sqrt{\frac{uK}{v_F}} =\frac{u'}{v_F}.
\end{equation}
This derivation used the long-wavelength part of the density, which can be expressed through a boson field $\Phi$: $\rho=-\partial_x\Phi$. Indeed, one can retrace the same steps by using the total density $\rho_{tot}$ both in the definition of the local potential (\ref{Vloc}) and in the interaction Hamiltonian kept in its initial form. Then Eq. (\ref{eqloc}) acquires an additional term: $V_{loc}\sin 2(\Phi-k_Fx)$. This adds nonlinear dependence of the conductance on $V_{loc}(2k_F)$ (even though this point needs more care). 
Thus the linear response to $E_{loc}$ in the presence of interactions with any range
is given by the external response of a wire with {\em short-range interactions}
parametrized by $u^{\prime },K^{\prime }$. Note that the backscattering process would affect $K'$ if it was not accounted for in the definition of the local potential, Eq. (\ref{Vloc}).\cite{these}\\ 
 Consider first the case where the product $uK$ is uniform all over the system, which is, for instance, the case in a homogeneous wire. Then,  Eq. (\ref{eqloc}) becomes a simple wave equation whose solution yields
\begin{equation}\label{jEfreq}
j(x,\omega )=K^{\prime }\int dye^{i\omega \left| x-y\right|
/u^{\prime }}E_{loc}(y,\omega). 
\end{equation}
It is worth inferring the frequency-dependent conductivity in response to the local field $i\omega\sigma'(\omega)=u'K'=uK$. Thus in the particular case of short-range interactions, the Drude peak height is not modified by the self-consistency of the potential. In order to find the conductance, we take the zero-frequency\cite{limit} limit of Eq. (\ref{jEfreq}):
$$j=g^{\prime }\int E_{loc}(y)=g^{\prime }\left[ V_{loc}(-\infty
)-V_{loc}(+\infty )\right]$$ with $g'=(2e^2/h) K'$ [Eq. (\ref{glocal})] once the charge and Planck's constant are restored.
   In a Galilean
invariant system, $uK=v_F$,\cite{haldanebosonisation} $g^{\prime}$ thus does not depend on the interactions.\cite{kawabata,oregrecent} In general, this is not true in the continuum limit of lattice systems. We can get the dominant effect of irrelevant umklapp process\cite{fujimoto_umklapp} or electron-phonon interactions by injecting the
renormalized charge stiffness found respectively in Ref.\onlinecite{giamarchi_rho} or Ref.\onlinecite{voit_phonon_recent} in Eq. (\ref{glocal}). In the latter case, we can go beyond the results of Ref.\onlinecite{brandes}. In the same way, one gets the effect on the conductance $uK/v_F$ obtained if one adopts the hypothesis of Ref.\onlinecite{egger}.

If $uK$ is not uniform over the wire, one has to solve a wave equation (\ref{eqloc}) with space-dependent parameters to find the response to $E_{loc}$.
In the geometry with leads, this again yields a conductance $g'=2e^2/h$ determined by $K'=1$ of the leads. The stationary current depends only on the asymptotic values of the potential that are not affected by the interactions absent on the leads: $V_{loc}(\pm \infty)=V(\pm \infty)$. Note that this holds also in the presence of impurities. But the alternative response is modified if self-consistency is taken into account. It is trivial if $uK=v_F$ in the wire: Eq. (\ref{jEfreq}) corresponds to the ac response of a noninteracting system.

Let us now discuss some conceptual problems.
 In Ref.\onlinecite{oregrecent}, it is argued that the
dissipation is determined by the local field.\ This is sensible, but
the experiments usually do not have access to the dissipative conductance. It would be possible to impose the current then measure the potential drop near the wire. This would yield an infinite conductance of the pure wire.\cite{landauer:70} Indeed, Apel and Rice\cite{apel_rice} suggested a similar four-probe measurement for the TLL (which turns out
to be not well defined due to a forgotten term in the potential drop%
\cite{these}). 
 Even in a multi-probe measurement, the possible invasive effect of a local potential probe leads to use the potential values imposed by the reservoirs\cite{buttickermulti} that determine alone the linear dc response. One has to worry about the
self-consistent field in two cases: the nonlinear regime or the ac
transport.\cite{landauer:70,buttickermulti,butticker_pretre} In a TLL, Eq. (\ref{jrho}) with the electric field $E$ as a source term shows that the current is exactly linear in $E$.\cite{ines,ines_nato} The nonlinearity appears if one considers the $2k_F$ momentum component, coupling between plasmons\cite{haldanebosonisation} or backscattering by impurities.\cite{egger} Concerning the ac regime, our analysis yields the dynamic conductivity as a response to the local field. Nevertheless, the ac response is more sensitive to experimental
setup details and to capacitive effects,\cite{butticker_pretre} and a more realistic model is required.

To summarize, we connect an interacting quantum wire to noninteracting leads in order to simulate the role of the reservoirs. The interactions can be of any range but have to be screened between the leads and the wire and to conserve momentum. The linear dc conductance does not depend on the interactions $g=2e^2/h$. We can either adopt Landauer's spirit (the reservoirs inject electrons that turn out to be perfectly transmitted) or we can include in the Hamiltonian the effect of an external electrostatic potential whose asymptotic values are again imposed by the reservoirs. In both alternatives, one does not have to know the field distribution through the structure, only the boundary conditions imposed by the reservoirs play a role. If case those are not fixed, the response to the local electrostatic potential is considered for general interactions. This yields a dc conductance $g'$ that depends on the charge stiffness, thus restricting the validity of the universal value $2e^2/h$ claimed in recent works to the situation where the current is not renormalized.

The author is grateful to H. J. Schulz and D. C. Glattli for stimulating discussions.

\begin{thebibliography}{10}
\bibitem{remarque}
While Ref.\onlinecite{ines} insisted more on the microscopic proof of Landauer's formula, here we reproduce the standard argument to stress the
  utility of the concept we proposed.
\bibitem{limit}
Again, the lengths on which the stationary regime holds are much less than
  $u/\omega$.
\end{thebibliography}

\end{document}